\documentclass[reprint, aps, prx, floatfix,superscriptaddress]{revtex4-2}
\usepackage{graphicx}    % needed for figures
\usepackage{dcolumn}    % needed for some tables
\usepackage{bm}             % for math
\usepackage{amssymb}   % for math
\usepackage{amsmath}    % for multiline equation
\usepackage{amsthm}
\usepackage{mathrsfs}    % for mathscr{}
\usepackage{commath}   % for absolute value
\usepackage{subfigure}    % for subfigure
\usepackage{braket}
\usepackage{color}
\usepackage{siunitx}
\usepackage[colorlinks=true, allcolors=blue]{hyperref}
\usepackage{hyphenat}
\usepackage{footnote}
\usepackage{notes2bib}
\usepackage{xcolor}
\newcommand{\beq}{\begin{equation}}
\newcommand{\eeq}{\end{equation}}

\begin{document}

\title{Capability of anti-degradable quantum channel with additional entanglement}

\author{Changchun Zhong}
\email{zhong.changchun@xjtu.edu.cn}

\affiliation{Department of Physics, Xi'an Jiaotong University, Xi'an, Shanxi 710049, China}
\affiliation{SeQure, Chicago, IL, 60615, USA}

\date{\today}

% \begin{abstract}

% \end{abstract}

\maketitle

\textbf{Quantum communication theory focuses on the study of quantum channels for transmitting quantum information, where the transmission rate is measured by quantum channel capacity. This quantity exhibits several intriguing properties, such as non-additivity, superactivation and so on. In this work, we show that a type of quantum channel known as the anti-degradable one-mode Gaussian channel—whose capacity is believed to be zero—can be ``activated" to transmit quantum information through the introduction of quantum entanglement. Although the channel's output alone cannot be used to retrieve the input signal, combining it with extra entanglement makes this possible. Beyond its theoretical implications, this activation can also be realized in practical systems. For example, in electro-optic systems used for quantum transduction in the two-mode squeezing interaction regime, the transduction channel is anti-degradable. We demonstrate that this system can transmit microwave-optical quantum information with the assistance of entanglement with an ancillary mode. This results in a new type of quantum transducer that exhibits positive quantum capacity over a wide parameter space.}

\textit{Introduction}--Quantum channel models the quantum information transmission through time or space. The study of various noisy quantum channels and their potential information transmission rate---quantum channel capacity---is central to quantum communication theory. Unlike its classical counterpart, quantum channel capacity does not have a simple formula, and its evaluation usually involves the so-called two-letter optimization which is computationally difficult \cite{shor2002,seth1997}. As a result, quantum capacity exhibits a series of unusual behaviors, such as activation and superactivation \cite{smith2008,smith2011,lim2018,lim2019}, which reflect the non-trivial way of quantum information traveling through the channel. The exact value of quantum capacity is known only for a few specific types of quantum channels. One such channel, known as anti-degradable, has been shown to have zero quantum capacity \cite{caruso2006,weedbrook2012}, meaning no quantum information can pass through the channel with vanishing error. In this paper, we show that a type of anti-degradable \textit{bosonic Gaussian channel} can have a non-vanishing quantum information transmission rate if the channel is combined with an ancillary mode. This is achieved by introducing initial squeezing and anti-squeezing between the ancillary mode and the mode involved in the anti-degradable Gaussian channel. These operations activate the channel's information transmission capability, which resembles superactivation phenomenon of quantum channels, although the protocol is quite different. The observation is against the common belief that anti-degradable channels are ineffective \cite{devetak2005}, and it suggests that some form of information is hidden in the channel output and can be used to recover the quantum information when properly combined with other signals. Furthermore, we show that this activation phenomenon occurs naturally in the realistic setting of quantum transduction with a two-mode squeezing interaction. This paves the way for a new design to realize quantum transduction for quantum networks.

\begin{figure}[t]
\centering
\includegraphics[width=\columnwidth]{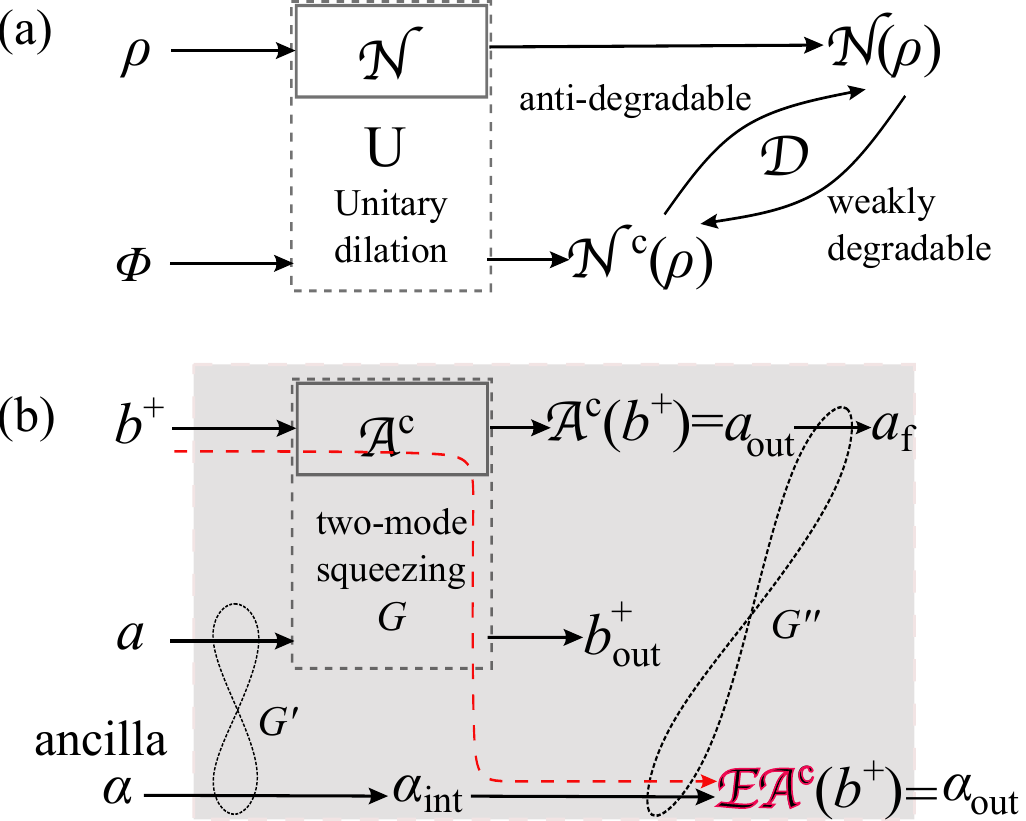}
\caption{(a) A quantum channel $\mathcal{N}$ comes with a unitary dilation, which defines a complementary channel $\mathcal{N}^c$. $\mathcal{N}$ is called degradable (anti-degradable) if there is a channel $\mathcal{D}$ to simulate the complementary output (if the other way round). (b) An anti-degradable channel $\mathcal{A}^c$ that is complementary to an amplification channel, which is generated by a two-mode squeezing interaction. By introducing an ancilla and mode entanglement, we can obtain an entanglement activated quantum channel that admits a non-zero quantum capacity $\textit{E}\mathcal{A}^c:\hat{b}^\dagger_{in}\rightarrow\textit{E}\mathcal{A}^c(\hat{b}^\dagger_{in})$, as indicated by the red dashed line with arrow. The grey block can be considered as a dilation of the channel $\textit{E}\mathcal{A}^c$. \label{fig1}}
\end{figure}

\textit{Bosonic Gaussian channel review}--As shown in Fig.~\ref{fig1}(a), a general bosonic quantum channel is represented by a completely positive and trace-preserving (CPTP) map, denoted as $\mathcal{N}:\rho\rightarrow\mathcal{N}(\rho)$. According to the dilation theorem, this channel can be considered as part of a unitary interaction involving ancillary input \cite{paulsen2002,wilde2013}, i.e., $\mathcal{N}(\rho)=tr_{anc}[U\rho\otimes{\Phi}_{anc} U^\dagger]$.
Note that, in this dilated channel representation, one can also trace out the system degrees of freedom to get a complementary output. This gives rise to a complementary channel $\mathcal{N}^c: \rho \rightarrow \mathcal{N}^c(\rho)$, where $\mathcal{N}^c(\rho) = \text{tr}{s}[U \rho \otimes \Phi_{anc} U^\dagger]$. For some specific quantum channels, the two outputs, $\mathcal{N}(\rho)$ and $\mathcal{N}^c(\rho)$, are related by a CPTP map $\mathcal{D}$. For example, if $\mathcal{N}^c(\rho) = \mathcal{D} \cdot \mathcal{N}(\rho)$, the channel $\mathcal{N}$ is referred to as \textit{weakly degradable}. Conversely, if $\mathcal{N}(\rho) = \mathcal{D} \cdot \mathcal{N}^c(\rho)$, the channel is termed \textit{anti-degradable}. Note that the definition of weakly degradability is reduced to \textit{degradability} if the ancillary input is confined to pure states \cite{caruso2006}.

A Gaussian channel is defined by a unitary interaction generated exclusively by a quadratic Hamiltonian acting on the bosonic modes, which guarantees a Gaussian output when the input state is Gaussian. It serves as a standard model for describing the noisy evolution of bosonic states and plays a crucial role in the theory and application of continuous-variable quantum information.

Since any n-mode Gaussian state $\rho(\mathbf{x,V})$ is fully determined by its first and second moments, with $\mathbf{x}=[x_1,p_1,x_2,p_2,...]^t$ and $\mathbf{V}$ the corresponding covariance matrix, the action of Gaussian quantum channels can be specified as \cite{weedbrook2012}
\begin{equation}
\mathbf{x}\rightarrow\mathbf{T}\cdot\mathbf{x}+\mathbf{d}, \mathbf{V}\rightarrow
\mathbf{T}\cdot\mathbf{V}\cdot\mathbf{T}^t+\mathbf{N},
\end{equation}
where $\mathbf{d}\in\mathcal{R}^n$ is a displacement vector and $\mathbf{T,N}$ are ${2n\times 2n}$ real matrices which satisfy the CPTP condition $\mathbf{N}+i\mathbf{\Omega}-i\mathbf{T\Omega T}^t\ge 0$ \cite{serafini2023,de2006}, where the matrix $\mathbf{\Omega}=\oplus^n\mathbf{\omega}$ is called the symplectic form with $\mathbf{\omega}=[(0,1),(-1,0)]$.

According to Holevo \cite{holevo2007}, a single-mode Gaussian channel($n=1$) admits a canonical classification, which is fully specified by the invariants $(\tau,r,\bar{n}_{e})$ defined by
\begin{equation}\label{eqref2}
\begin{split}
    &\tau=\text{det}(\mathbf{T}),r=\text{min}\{\text{rank}(\mathbf{T}),\text{rank}(\mathbf{N})\},\\
    &\bar{n}_{e}=\begin{cases}
        \sqrt{\text{det}(\mathbf{N})}, &\tau=1 \\ 
        \frac{\sqrt{\text{Det}(\mathbf{N})}}{2\abs{1-\tau}}-\frac{1}{2},&\tau\neq 1
    \end{cases}.
\end{split}
\end{equation}
The case with $0<\tau<1$ ($\tau>1,\tau=1$) corresponds to the well-known bosonic loss (amplification, random displacement) channel. Specifically, $\tau<0$ defines a class of one-mode Gaussian channels, which are known to be anti-degradable \cite{caruso2006,holevo2007}. For these Gaussian channels, their quantum channel capacity are lower bounded by the following formula
\begin{equation}\label{eqqlb}
Q_\text{LB}=
\begin{cases}
      \max\{0,\log_2|\frac{\tau}{1-\tau}|-g(\bar{n}_e)\}, & \tau\neq 1 \\
      \max\{0,\log_2(\frac{2}{e\sigma^2})\},& \tau=1, 
\end{cases}
\end{equation}
where $g(\bar{n}_e)=(\bar{n}_e+1)\log_2(\bar{n}_e+1)-\bar{n}_e\log_2\bar{n}_e$ and $\sigma^2=\sqrt{\text{det}\mathbf{N}}$.

\textit{The amplification channel and its complement}--In this paper, we focus on the less-investigated channel with $\tau<0$. This channel is sometimes called a phase-conjugation channel, which can be obtained by taking the complement to an amplification channel, as schematically shown in Fig.~\ref{fig1}(b). For illustration, we first write down the amplification channel $\mathcal{A}$ in terms of the mode operators
\begin{equation}\label{eqref4}    \mathcal{A}:\hat{b}\rightarrow\hat{b}_\text{out}=\sqrt{G}\hat{b}+\sqrt{G-1}\hat{a}^\dagger,
\end{equation}
where $G>1$ is the amplification coefficient. The complementary channel is thus
\begin{equation}\label{eqref1}
\mathcal{A}^c:\hat{b}^\dagger\rightarrow\hat{a}_\text{out}=\sqrt{G-1}\hat{b}^\dagger+\sqrt{G}\hat{a}
\end{equation}
This channel $\mathcal{A}^c$ can be further put in the quadrature representation $\mathbf{x}_{out}=\mathbf{T}\cdot\mathbf{x}_{in}+\mathbf{d}$ as
\begin{equation}
    \begin{pmatrix}
        x^a_{out}\\p^a_{out}
    \end{pmatrix}=
    \begin{pmatrix}
        \sqrt{G-1} & 0 \\
        0 & -\sqrt{G-1}
    \end{pmatrix}
    \begin{pmatrix}
        x^b \\ p^b
    \end{pmatrix}+
    \begin{pmatrix}
        \sqrt{G}x^a \\ \sqrt{G}p^a
    \end{pmatrix},
\end{equation}
where the quadratures and the corresponding mode operators are related by $x=\hat{a}+\hat{a}^\dagger,p=i(\hat{a}^\dagger-\hat{a})$. Straightforwardly, we have $\tau=-(G-1)<1$, which falls into the last category according to Holove's canonical classification of one-mode Gaussian channels \cite{holevo2007}. The channel $\mathcal{A}^c$ is known to be anti-degradable, meaning it has zero quantum capacity. This can be intuitively understood because the input $\hat{a}$ (consider it as noise input in this context) always has a larger coefficient than the signal input $\hat{b}^\dagger$ due to $G>1$. In the following section, we will show that a non-zero quantum capacity can be achieved by introducing entanglement with an ancillary mode.

\textit{Quantum capacity activation with entanglement}--As mentioned earlier, the channel $\mathcal{A}^c$ is the complement of an amplification channel, where the corresponding unitary dilation is known as a two-mode squeezing operator $S(G)=\exp[r(\hat{a}\hat{b}+\hat{a}^\dagger\hat{b}^\dagger)/2]$, with $G=\cosh r$ and $r$ is the squeezing factor. If the initial states are both vacuum, the operator generates a two-mode squeezed vacuum state with entanglement between the modes $\hat{a}$ and $\hat{b}$ with each mode containing $\sinh^2r$ photons. To potentially activate the quantum capacity of $\mathcal{A}^c$, we introduce an ancillary mode $\hat{\alpha}$ and entangle it with the noise input mode $\hat{a}$ using a two-mode squeezing operation $S(G^\prime)$, as shown in Fig.~\ref{fig1}(b). A second two-mode anti-squeezing operation $S^\dagger(G^{\prime\prime})$ is applied between the ancilla and the output of the complement channel $\mathcal{A}^c$. Mathematically, these operations are equivalent to the mode operators undergoing the following transformations: For the first entangling operation:
\begin{equation}
\begin{split}
\hat{a}&\rightarrow\sqrt{G^\prime}\hat{a}+\sqrt{G^\prime-1}\hat{\alpha}^{\dagger}\\
\hat{\alpha}&\rightarrow\sqrt{G^\prime-1}\hat{a}^\dagger+\sqrt{G^\prime}\hat{\alpha}=\hat{\alpha}_{int}.
\end{split}
\end{equation}
For the second anti-squeezing operation:
\begin{equation}
\begin{split}
\hat{a}_{out}&\rightarrow\sqrt{G^{\prime\prime}}\hat{a}_{out}-\sqrt{G^{\prime\prime}-1}\hat{\alpha}_{int}^{\dagger}=\hat{a}_f\\
\hat{\alpha}_{int}&\rightarrow-\sqrt{G^{\prime\prime}-1}\hat{a}^\dagger_{out}+\sqrt{G^{\prime\prime}}\hat{\alpha}_{int}=\hat{\alpha}_{out}.
\end{split}
\end{equation}
The intuition behind these operations is simple: we try to reduce or cancel the noise in the channel output $\mathcal{A}^c(\hat{b}^\dagger)$. As seen in Eq.~\ref{eqref1}, the noise originates from the mode $\hat{a}$. Therefore, the first entangling operation allows the ancillary mode $\hat{\alpha}$ to carry the noise information, while the second anti-squeezing serves to cancel out this noise in the output \cite{shi2024o}. 

However, this intuition is not immediately correct, which can be seen by writing down the channel from $\hat{b}$ to $\hat{a}_f$ as the transformation $\mathbf{x}^{{a}_f}=\mathbf{T}\cdot\mathbf{x}^{{b}}+\mathbf{d}$, where $\hat{a}_f$ denotes the amplified mode of the second squeezing and $\mathbf{x}^{(a_f,b)}$ labels the quadratures of the corresponding modes. The matrix $\mathbf{T}=[(\sqrt{G^{\prime\prime}(G-1)} , 0),(0, -\sqrt{G^{\prime\prime}(G-1)})]$. Obviously, we have det$(\mathbf{T})=-G^{\prime\prime}(G-1)<0$, which indicates that it is still a phase conjugation channel (anti-degradable) with zero quantum capacity. This result is contradictory to the belief that the operations we introduce should be helpful in some way. This naturally leads us to check the channel from $\hat{b}$ to $\hat{\alpha}_{out}$, where $\hat{\alpha}_{out}$ is the other output mode of the second anti-squeezing operation. Similarly, we write down this channel in the quadrature representation
\begin{equation}
\begin{split}
\mathbf{x}^{\alpha_{out}}&=\mathbf{T}\cdot\mathbf{x}^b+\mathbf{d}\\
\mathbf{V}^{\alpha_{out}}&=\mathbf{T}\cdot\mathbf{V}^b\cdot\mathbf{T}^t+\mathbf{N}
\end{split}
\end{equation}
where
\begin{equation}
    \mathbf{T}=
    \begin{pmatrix}
        -\sqrt{G^{\prime\prime}-1}\sqrt{G-1} & 0 \\
        0 & -\sqrt{G^{\prime\prime}-1}\sqrt{G-1}
    \end{pmatrix},
\end{equation}
and the matrix $\mathbf{N}=[(m,0),(0,m)]$ with
\begin{equation}\label{eqref3}
\begin{split}
m=&(\sqrt{G^\prime} \sqrt{G^{\prime\prime}} - \sqrt{G^\prime-1} \sqrt{G^{\prime\prime}-1} \sqrt{G})^2 \\  &+(\sqrt{G^\prime-1} \sqrt{
    G^{\prime\prime}} - \sqrt{G^\prime} \sqrt{G^{\prime\prime}-1} \sqrt{G})^2.
\end{split}
\end{equation}
Remarkably, this channel admits a non-zero quantum capacity if we appropriately tune the parameters. To catch the physical meaning, we call this channel \textit{the entanglement-activated anti-degradable channel}, denoted by 
\begin{equation}\label{eqrefeac}
\textit{E}\mathcal{A}^c:\hat{b}\rightarrow\hat{\alpha}_{out}.
\end{equation}
This channel has a channel transmissivity $\tau=(G-1)(G^{\prime\prime}-1)$ which is positive, and thus defines the three types of bosonic Gaussian channels depending on the values of $G$ and $G^{\prime\prime}$ (loss, amplify and random displacement channel). The corresponding channel noise can be obtained by combining Eq.~\ref{eqref2} and Eq.~\ref{eqref3}, which is a general function of all the squeezing coefficients and can approach zero when the squeezing strengths are properly chosen.  

\begin{figure}[t]
\centering
\includegraphics[width=\columnwidth]{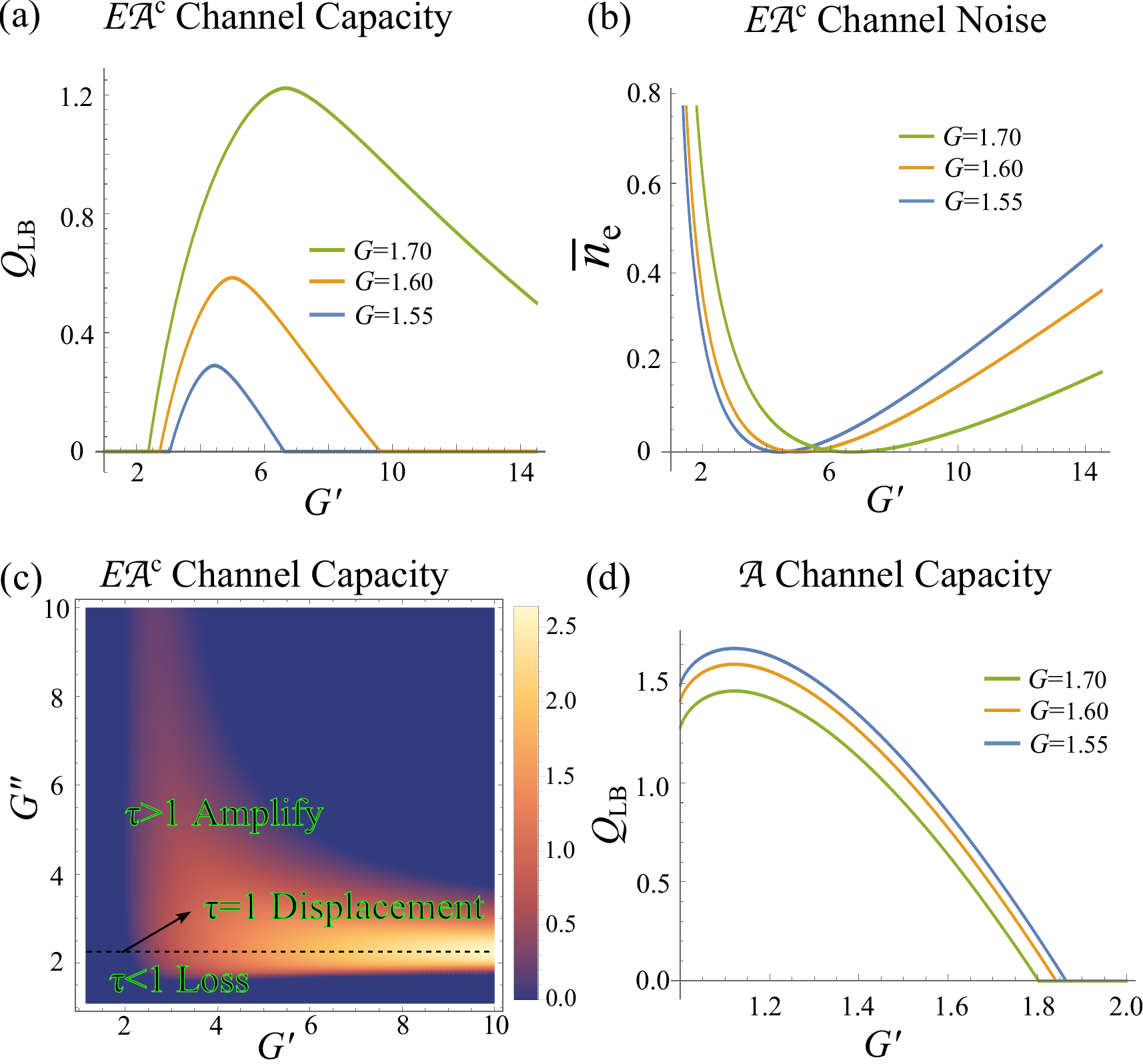}
\caption{(a) The quantum capacity lower bound in terms of the initial squeezing $G^\prime$ with fixed $G^{\prime\prime}=2$. (b) The channel noise in terms of the initial squeezing $G^\prime$ with fixed $G^{\prime\prime}=2$. (c) The quantum capacity lower bound with respect to $G^\prime$ and $G^{\prime\prime}$, fixing $G=1.8$. The dashed black line traces the parameter $G^{\prime\prime}=2.25$ such that $\tau=1$, giving an effective random displacement channel. The area below and above the line correspond to an effective bosonic loss channel and amplification channel, respectively. (d) The capacity lower bound for the channel $\mathcal{A}$ in terms of $G^\prime$. \label{fig2}}
\end{figure}

To better show the channel activation, we numerically evaluate the lower bound of the quantum capacity using Eq.~\ref{eqqlb}. In Fig.~\ref{fig2}(a), we calculate the lower bound in terms of the initial squeezing $G^\prime$ with the second squeezing fixed at $G^{\prime\prime}=2$. The three curves, corresponding to different amplification factors
$G$, clearly demonstrate the positive capacity. Figure~\ref{fig2}(b) shows the channel noise corresponding to all the parameters depicted in Fig.~\ref{fig2}(a). Interestingly, the noise can be totally canceled by tuning $G^\prime$, and the minimal points correspond to those peaks of the curves for the quantum capacity, as depicted in Fig.~\ref{fig2}(a). Figure~\ref{fig2}(c) scans the parameter regime ($G^\prime,G^{\prime\prime}$) that admits positive quantum capacity lower bound with $G=1.8$. We see the positive area spreads to almost all the lower half of the plane. Since $G=1.8$ is chosen, the channel transmissivity $\tau=1$ is obtained at $G^{\prime\prime}=2.25$. It is indicated by the black dashed line, which defines the effective random displacement channel and the area below (above) gives the effective bosonic loss (amplification) channel. 

It is worth noting that there is a concern about violating the no-cloning theorem \cite{wootters2009}, since the original amplification channel $\mathcal{A}:\hat{b}\rightarrow\hat{b}_{out}$ defined in Eq.~\ref{eqref4} might have a non-zero quantum capacity. Indeed, it is a thermal amplification channel with thermal noise determined by the initial squeezing $\bar{n}_{e}=G^\prime-1$. As shown in Fig.~\ref{fig2}(d), the curves depicts the capacity lower bound of $\mathcal{A}$. Choosing the same parameter $G$ as in Fig.~\ref{fig2}(a), the quantum capacity is positive for small $G^\prime$ but safely drops to zero when tuning up $G^\prime$ (thermal noise $\bar{n}_{e}$ increases), before the channel $\textit{E}\mathcal{A}^c$ starting to have non-zero quantum capacity. This suggests that the no-cloning theorem is safe.

The significance of this finding is twofold: 1)
the anti-degradable channel having zero quantum capacity is activated with the assistance of quantum entanglement, which reveals the non-trivial quantum information flow in the channels; 2) if we encode information in the $\hat{b}$ mode and send it through the channel, the quantum information (with non-zero channel capacity) can be recovered in the ancillary mode. This process resembles the well-known phenomenon of quantum teleportation, but without the need for classical measurements or feedback operations, which could be useful in many practical settings. The practical implications of this scheme will become clearer in the next section.

\textit{Application to quantum transducer}--Quantum transducer aims to convert quantum information between microwave and optical photons \cite{Andrews2014,zhong2020}, which plays an essential role in modern quantum technology, such as large scale quantum networks \cite{cirac1997,kimble2008}, and its realization is long hindered by certain limitations including low transduction transmissivity, excessive added noise and narrow bandwidth \cite{han2021,zhong2024,lauk2020,meesala2024}. Designing an efficient transducer is thus a very active research topic in the quantum information community \cite{Regal2011,Bochmann2013,Taylor2011,Barzanjeh2012,Wang2012,Tian2010,*Tian2012,*Tian2014,Zou2016,Midolo2018,Bagci2014,Vainsencher16,Winger2011,Pitanti2015,mayor2024,zhao2024,wu2021,stefan2021,han2020ca,meesala2023,Tsang2010,Tsang2011,Javerzac-Galy2016,fan2018,zhong2022PRAPP,Hisatomi2016,Hafezi2012,Kiffner2016,Gard2017,zhu2020}. In this section, we show that the channel $\textit{E}\mathcal{A}^c$ naturally defines a new type of quantum transducer.

Without loss of generality, we take an electro-optic (EO) system based quantum transducer for example. This system involves a three-wave mixing interaction among two optical cavity modes and a microwave superconducting resonator enabled by certain material with Pockels nonlinearity \cite{fan2018,Tsang2010,Tsang2011}. Using a blue detuned optical drive, the three wave mixing interaction usually reduces to a two-mode-squeezing coupling {with rotating wave approximation}
\begin{equation}    \hat{H}/\hbar=g\hat{a}\hat{b}+g^*\hat{a}^\dagger\hat{b}^\dagger,
\end{equation}
where we denote $\hat{a}$ and $\hat{b}$ as the optical and microwave mode operators, and $g$ is the mode coupling strength. This Hamiltonian is capable of generating microwave-optical entanglement, which is the basis for the \textit{entanglement-based quantum transducer}---using the entanglement to teleport quantum information between the two frequencies \cite{zhong2020,zhong2020pra}. Since the squeezing coupling is parametrically generating photons, the input microwave (optical) signal will be amplified, which is exactly the amplification channel defined in Eq.~\ref{eqref4}. The transduction channel, e.g., from microwave to optical state, is the complementary channel as defined in Eq.~\ref{eqref1}. In the context of EO system, the parameter $G$ can be obtained by solving the Heisenberg-Langevin equation in the frequency domain and combining the input-output relationship \cite{Tsang2011,zhong2022PRR,fu2021}. With the system on resonance, we obtain
\begin{equation}
    G=\left(\frac{1+C_g}{1-C_g}\right)^2,
\end{equation}
where $C_g=4\abs{g}^2/\kappa_o\kappa_e$ is the cooperativity of the EO system, and $\kappa_o$ ($\kappa_e$) are the optical (microwave) mode dissipation rate. Note that it is necessary to have $C_g<1$ for the system to be stable. In the state-of-the-art experiment, $C_g$ is still much less than one, satisfying the stability condition \cite{fu2021}. 

\begin{figure}[t]
\centering
\includegraphics[width=\columnwidth]{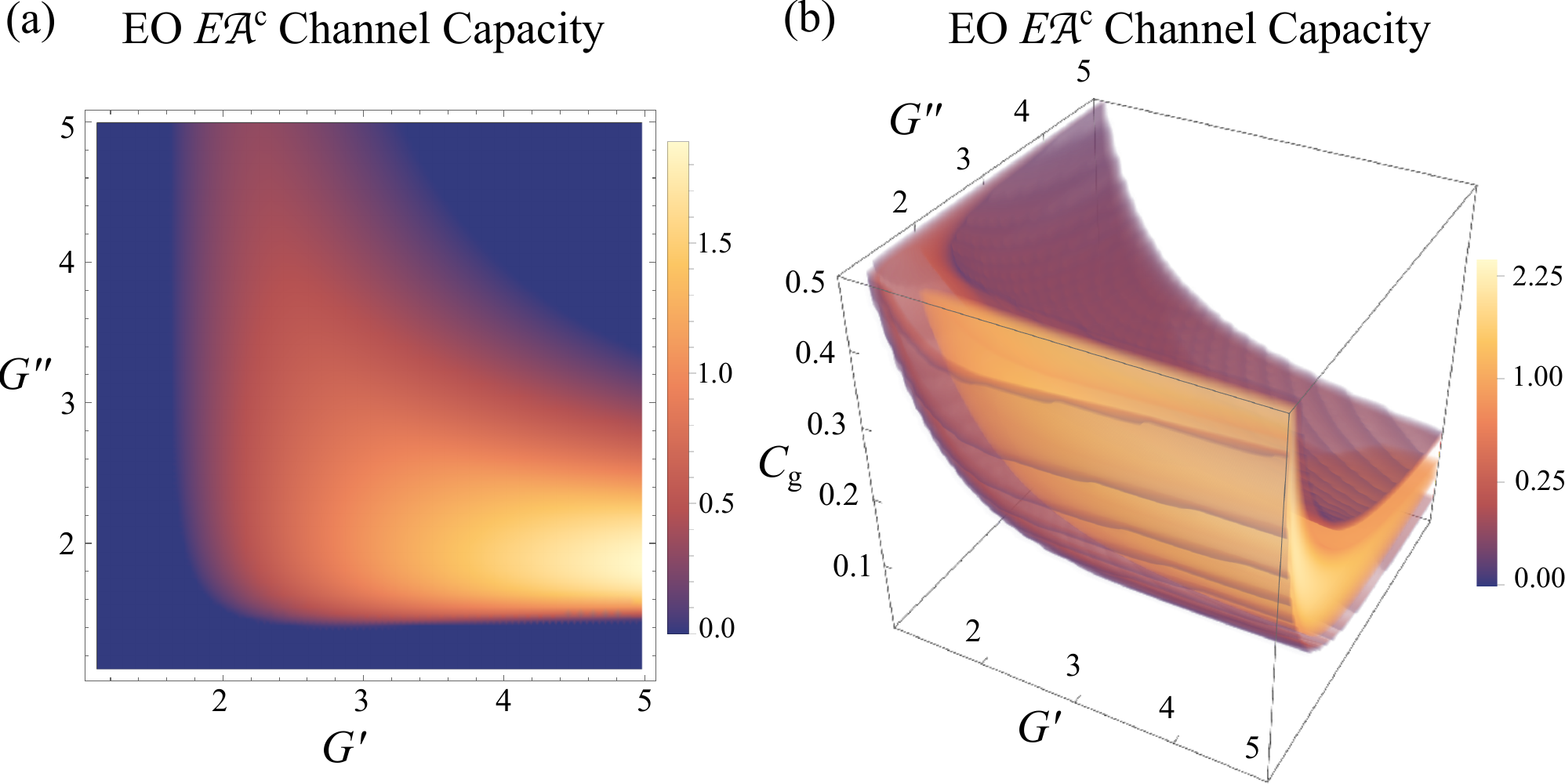}
\caption{(a) The quantum capacity lower bound of EO-based $\textit{E}\mathcal{A}^c$ channel in terms of the squeezing $G^\prime$ and $G^{\prime\prime}$ for the system cooperativity $C_g=0.1$. (b) The quantum capacity lower bound of EO-based $\textit{E}\mathcal{A}^c$ channel in terms of the squeezing $G^\prime$, $G^{\prime\prime}$ and $C_g$.\label{fig3}}
\end{figure}

Since this transduction channel for any $C_g$ is anti-degradable, its potential for quantum transduction has long been ignored. As discussed previously, we can introduce one ancillary mode to activate its potential. Depending on the transduction direction, the ancillary mode should have a working frequency either in the optical or microwave regime. Taking the microwave-to-optical transduction for illustration, as shown in Fig.~\ref{fig1}(b), the ancilla $\hat{\alpha}$ is an optical mode, which can couple to the optical $\hat{a}$ mode using a two-mode squeezer with both modes in the optical domain. The microwave signal denoted by $\hat{b}^\dagger$ is injected to the EO transducer where we take the converted signal $\hat{a}_{out}$ and couple it with the ancilla using another two-mode squeezer. Note that this squeezer also only involves modes in the optical domain, thus can be straightforwardly implemented in experiment. The only non-trivial part is the EO system itself which involves coupling between microwave and optical modes. As demonstrated earlier, the final output of the ancilla (optical mode) can recover the information from the microwave input. This is evidenced by the non-zero quantum capacity of the entanglement-activated anti-degradable channel $\textit{E}\mathcal{A}^c$ defined in Eq.~\ref{eqrefeac}. 

To explicitly show that, we calculate the quantum capacity lower bound of the EO-based $\textit{E}\mathcal{A}^c$ channel. As shown in Fig.~\ref{fig3}(a), by fixing the system cooperativity $C_g=0.1$, a large region in the parameter space $(G^\prime,G^{\prime\prime})$ exhibits positive channel capacity. The transduction channel is activated and, in general, enhanced by properly tuning the squeezing factors. The region with zero capacity lower bound corresponds to excessive channel noise, as indicated by the schematics in Fig.~\ref{fig2}(b). Figure~\ref{fig3}(b) presents a 3D density plot where we additionally scan the system cooperativity. A larger $C_g$ tends to yield positive quantum capacity with much smaller squeezing factors $G^\prime$ and $G^{\prime\prime}$. In experimental design, it is advisable to optimize these parameters according to practical limitations.

It is worth noting that, for an optical-to-microwave transducer, an microwave ancilla is needed such that we can apply the two-mode squeezer in the microwave domain. Thus, in order to have a bi-direction quantum transducer, we would need two ancillary modes, one with working frequency in the optical regime and the other in microwave. Depending on the task, we can conveniently choose between them.

\textit{Discussion}--As a bizarre phenomenon, super-activation is conjectured to be related to the fact that different channels might carry different types of quantum information that enhance each other when combined \cite{smith2008,smith2011}. Although the protocol is quite different, the ``activation" discussed in our context strongly indicates a similar conjecture. The output from the anti-degradable channel hides the signal (i.e., the hidden information), while the ancillary mode carries the necessary entanglement, and the mixing of them by squeezing interaction leads to a positive capacity of the channel $\textit{E}\mathcal{A}^c$.

Note that the anti-degradable channel can be simulated by its complement along with another CPTP map. This implies that at least two copies of the hidden information may be accessible. This cloning of the hidden information reveals its partial-classical nature. More interestingly, this raises a question about the relationship between no-clonability and quantum properties: specifically, how ``quantum" must a state be before it becomes unclonable? Is there a phase transition between states that are clonable and those that are not?

In practical applications of this protocol, a quantum transducer must account for physical limitations. A transducer based on any physical system will inevitably suffer from non-ideal dissipation or additional thermal noise. For instance, in the EO system, coupling to the mK-scale environment is intrinsic. This will degrade the transduction channel. Therefore, the system parameters, along with the additional squeezing factors, must be optimized during the experimental design. We leave a more detailed discussion of this for future work.

\begin{acknowledgments}
C.Z. thanks the start-up support from Xi'an Jiaotong University (Grant No. 11301224010717). C.Z. thanks his mom, who passed away a year ago, for her whole-hearted support.
\end{acknowledgments}

% \nocite{*}
\bibliography{all}

\onecolumngrid

\begin{appendix}

\end{appendix}

\end{document}